\documentclass{article}
\usepackage{spconf,amsmath,graphicx,multirow}
\usepackage{bbding}
\usepackage{booktabs}
\usepackage{cite}
\usepackage{verbatim}
\usepackage{bibspacing}
\usepackage[utf8]{inputenc}
\usepackage[titletoc,title]{appendix}
\usepackage{makecell}
\usepackage{tipx}

\newcommand{\xmark}{\ding{55}}

\usepackage{color}
\usepackage{multirow}
\usepackage{hhline}
\usepackage{subfigure}
\usepackage{pifont}
\usepackage{bm}
\usepackage{amsmath,amsfonts,amssymb,mathtools}

\usepackage{graphicx,float}

\title{Exploring Self-supervised Pre-trained ASR Models For Dysarthric and Elderly Speech Recognition}
\name{SJ Hu$^1$, XR Xie$^2$, ZR Jin$^1$, MZ Geng$^1$, Y Wang$^1$, MY Cui$^1$, JJ Deng$^1$, Xunying Liu$^1$, Helen Meng$^1$}
\address{
    $^1$The Chinese University of Hong Kong, Hong Kong SAR, China\\
  $^2$Institute of Software, Chinese Academy of Sciences, China
}

\begin{document}
\setlength{\bibitemsep}{.06\baselineskip}
\ninept
\maketitle

\begin{abstract}
Automatic recognition of disordered and elderly speech remains a highly challenging task to date due to the difficulty in collecting such data in large quantities. This paper explores a series of approaches to integrate domain adapted Self-Supervised Learning (SSL) pre-trained models into TDNN and Conformer ASR systems for dysarthric and elderly speech recognition: a) input feature fusion between standard acoustic frontends and domain adapted wav2vec2.0 speech representations; b) frame-level joint decoding of TDNN systems separately trained using standard acoustic features alone and with additional wav2vec2.0 features; and c) multi-pass decoding involving the TDNN/Conformer system outputs to be rescored using domain adapted wav2vec2.0 models. In addition, domain adapted wav2vec2.0 representations are utilized in acoustic-to-articulatory (A2A) inversion to construct multi-modal dysarthric and elderly speech recognition systems. Experiments conducted on the UASpeech dysarthric and DementiaBank Pitt elderly speech corpora suggest TDNN and Conformer ASR systems integrated domain adapted wav2vec2.0 models consistently outperform the standalone wav2vec2.0 models by statistically significant WER reductions of 8.22\% and 3.43\% absolute (26.71\% and 15.88\% relative) on the two tasks respectively. The lowest published WERs of 22.56\% (52.53\% on very low intelligibility, 39.09\% on unseen words) and 18.17\% are obtained on the UASpeech test set of 16 dysarthric speakers, and the DementiaBank Pitt test set respectively. 
\end{abstract}

\begin{figure*}[htbp]
    \centering
    \includegraphics[width=0.8\textwidth]{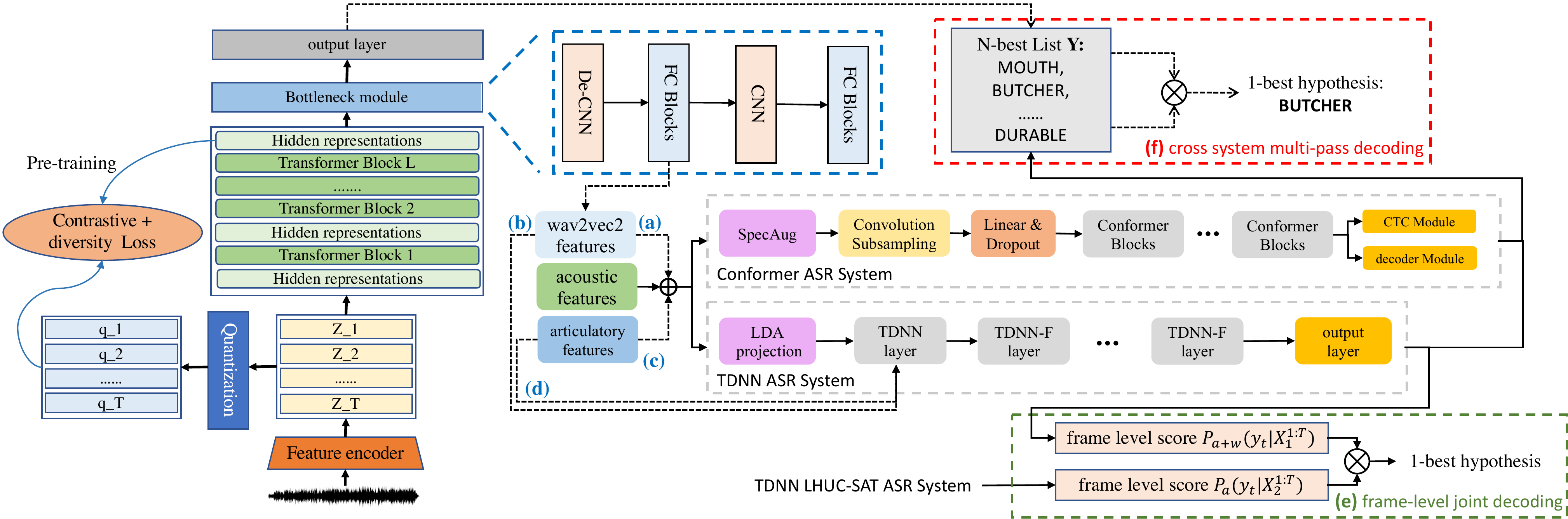}
    \vspace{-0.4cm}
    \caption{
    Example wav2vec2.0 model containing a ``Bottleneck module'' to extract domain adapted speech features (circled in blue dotted line) that are integrated into TDNN/Conformer ASR systems using: 1) input feature fusion with standard acoustic frontends via connections (a) and (b); 2) TDNN system frame-level joint decoding in green box (e); and 3) TDNN/Conformer $\rightarrow$ wav2vec2.0 multi-pass rescoring in red box (f). Connections (c) and (d) produce AASR systems using additional articulatory features predicted from wav2vec2.0 speech representations.
    }
    \label{fig:my_model}
    \vspace{-0.55cm}
\end{figure*}
\vspace{-0.4cm}
\section{Introduction}
\vspace{-0.3cm}
Accurate recognition of elderly and disordered speech remains a challenging task \cite{yu2018development,xiong2020source,liu2021recent,ye2021development,deng21d_interspeech,geng2022speaker, yue2022acoustic}.
Speech disorders such as dysarthria are associated with neuro-motor conditions
and often found among elderly adults experiencing neurocognitive disorders such as Alzheimer’s disease (AD). ASR technologies tailored for their needs not only improve their quality of life, but also support large scale automatic early diagnosis of neurocognitive impairment 
\cite{ferri2005global,vipperla2010ageing,rudzicz2014speech,zhou2016speech,mirheidari2019dementia,ye2021development}.
\par
Elderly and disordered speech bring many challenges to current deep learning based ASR technologies predominantly targeting non-aged, healthy adult users. First, a large mismatch between such data and non-aged, healthy adult voice is often observed.
Second, the co-occurring disabilities and mobility limitations often found among elderly and disordered speakers lead to the difficulty in collecting large quantities of such data that are essential for data intensive ASR system development. To this end, a series of recent researches have been conducted to transfer knowledge from normal speech and ASR systems trained on such out-of-domain data \cite{geng2022speaker, ye2021development, jin2022personalized, xie21b_interspeech, bhat22_interspeech, takashima2020two,yu2018development,xiong2020source,liu2021recent,deng21d_interspeech, wang2021improved, green21_interspeech, tobin2022personalized, wang2022conformer}. 

\par
An alternative solution to address the above data scarcity problem is to use self-supervised learning (SSL) models \cite{baevski2020wav2vec, chen2022wavlm, hsu2021hubert} that are pre-trained on large quantities of unlabelled data. The associated neural speech representations produced by these pre-trained ASR systems are also inherently robust to domain mismatch \cite{hernandez2022cross, hsu2021robust, 9688260}. Although they have been successfully applied to a range of normal speech processing tasks including speech recognition \cite{baevski2020wav2vec, conneau2020unsupervised, chen2022wavlm, hsu2021hubert}, speech emotion recognition \cite{pepino2021emotion} and speaker recognition \cite{vaessen2022fine}, very limited researches on SSL pre-trained models for disordered and elderly speech have been conducted \cite{violeta2022investigating, baskar2022speaker, hernandez2022cross}.
Among these, wav2vec2.0 and WavLM models were applied to Japanese electrolaryngeal and English dysarthric speech in \cite{violeta2022investigating}, where an overall WER of 51.8\% was reported on the benchmark UASpeech task. Speaker adaptation of wav2vec2.0 models using fMLLR and x-vectors for dysarthric speech recognition was investigated in \cite{baskar2022speaker}. Cross-lingual SSL based dysarthric speech recognition was studied in \cite{hernandez2022cross}.

\par
Efforts on applying SSL pre-trained speech models to dysarthric and elderly speech are also confronted with the same data scarcity problem. Direct fine-tuning of a large number of model parameters on limited impaired or elderly speech data rapidly leads to overfitting and poor generalisation. This issue is further exasperated when the limited training data provides insufficient coverage of words in the test data. For example, approximately 39\% of words in the benchmark UASpeech test set do not occur in the training data. This was found to produce a larger disparity in performance between seen and unseen words in dysarthric speech, and between impaired speakers with high and very low intelligibility, for end-to-end ASR systems including normal speech pre-trained models, than hybrid ASR systems that are  trained using phonetic targets \cite{liu2021recent,wang2021improved, violeta2022investigating}.
\par
To this end, this paper explores a series of approaches to integrate SSL pre-trained wav2vec2.0 models into hybrid TDNN \cite{peddinti2015time} and Conformer \cite{gulati2020conformer} ASR systems for dysarthric and  elderly speech recognition in order to exploit the diversity and complementarity among them, including
a) input feature fusion between standard acoustic front-ends and wav2vec2.0 speech representations; b) frame-level joint decoding \cite{swietojanski2013revisiting, liu2021recent, hu2022exploitinguti} of TDNN systems separately trained using standard acoustic features alone and with additional wav2vec2.0 representations; and c) cross system multi-pass decoding \cite{cui2022two} involving the TDNN or Conformer system N-best outputs to be rescored using domain adapted wav2vec2.0 models. Finally, wav2vec2.0 representations are utilized to facilitate neural network based acoustic-to-articulatory (A2A) inversion \cite{hu2022exploitinguti}. The resulting articulatory movement features are used to construct multi-modal dysarthric and elderly ASR systems.
\par
Experiments conducted on the UASpeech \cite{kim2008dysarthric} dysarthric and DementiaBank Pitt \cite{becker1994natural} elderly speech corpora suggest TDNN and Conformer ASR systems constructed by integrating domain adapted wav2vec2.0 models consistently outperform the standalone wav2vec2.0 models by statistically significant WER reductions of 8.22\% and 3.43\% absolute (26.71\% and 15.88\% relative) on the two tasks respectively. 
The lowest published WERs of 22.56\% (52.53\% on very low intelligibility, 39.09\% on unseen words) and 18.17\% are obtained 
on the UASpeech test set of 16 dysarthric speakers and the DementiaBank Pitt test set. 


\par
The main contributions of the paper are summarized below:\\ 
\textbf{1)} To the best of our knowledge, this paper presents the first work to systematically investigate the generalization capability of SSL pre-trained models for dysarthric and elderly speech recognition tasks including two objectives: a) minimize the overall WER with a focus on the most challenging portions of the very low (VL) intelligibility dysarthric or elderly speech data; and b) model generalization performance on unseen words that are unlikely to be sufficiently covered in practice for pathological and medical domain speech. In contrast, related previous works \cite{violeta2022investigating, baskar2022speaker, hernandez2022cross} analyzed their performance disparity across different speech pathology severity subsets but not their generalization to unseen words or sentences. \\
\textbf{2)} To address such generalization issues, this paper proposes novel approaches to integrate SSL pre-trained models into in-domain data constructed ASR systems. 
In contrast, prior researches focused on using stand-alone domain adapted pre-trained models \cite{violeta2022investigating, baskar2022speaker, hernandez2022cross}. \\
\textbf{3)} This work also presents the first use of SSL based speech representations to cross-domain A2A inversion. In contrast, prior researches on cross-domain A2A inversion were conducted predominantly using non-SSL based domain adaptation methods \cite{hu2022exploiting, hu2022exploitinguti}.
\vspace{-0.6cm}
\section{Pre-trained Wav2vec2.0 ASR Model}
\vspace{-0.2cm}
\subsection{Model Architecture}
\vspace{-0.2cm}
The wav2vec2.0 model is an SSL pre-trained model that jointly learns latent contextualized speech representations and an inventory of discretized speech units. It consists of three components (left part of Fig. 1), including a \textbf{speech feature encoder $f(\cdot)$}
containing multi-layer 1D CNN based convolution blocks, a \textbf{contextual transformer network $g(\cdot)$} and a \textbf{quantization module $h(\cdot)$} consisting of 2 codebooks with 320 entries each. The input of wav2vec2.0 model is raw speech audio $\mathcal{X}$, which is first encoded into a sequence of representations $\mathcal{Z}=f(\mathcal{X})$ with a stride of 20 ms and a receptive field of 25 ms by the CNN-based speech feature encoder. These CNN extracted features are then fed into transformer based contextual encoder to produce contextualized speech representations $\mathcal{C}=g(\mathcal{Z})$. A logit $l \in \mathcal{R}^{2 \times 320}$ is calculated from each vector $z_i \in \mathcal{Z}$ before fed into Gumbel-Softmax \cite{jang2016categorical} operators to choose the best code among all entries of each codebook. The selected codes are concatenated together before a linear transformation is applied to produce the quantized vector $q_i$ during pretraining.
\vspace{-0.6cm}
\subsection{Model Pre-training And Fine-tuning}
\vspace{-0.2cm}
Wav2vec2.0 model is pre-trained using a weighted interpolation between the contrastive and diversity loss functions. For the speech representation of time step $t$ that is randomly masked, the contrastive loss based training aims to reproduce the quantized vector $ q_t$ for the masked frame at the output of contextual transformer encoder, while discriminating against $K$ ``distractors''  representations ${\tilde q} \in {\tilde Q}$ that are randomly sampled from the other masked time steps within the same speech utterance. The contrastive loss function is defined as
\begin{small}
\begin{equation}
    \setlength{\abovedisplayskip}{2pt}
    \setlength{\belowdisplayskip}{2pt}
    L_{m} = - \log \frac{\exp({\rm sim}(c_t, q_t)/ \kappa)}{\sum_{\widetilde q \in \widetilde Q} \exp({\rm sim}(c_t, \widetilde q)/ \kappa)}
\end{equation}
\end{small}
\par
\noindent
where ${\rm sim}(c_t, q_t)$ is the cosine similarity between the contextualized vector $c_t$ and the quantized representation $q_t$. $\kappa$ is the temperature and set $\widetilde Q$ consists of both the target quantization $q_t$ and $K$ distractors. The entropy based diversity loss is designed to regularize the use of the quantized codebook representations.
\par
During the fine-tuning stage, a randomly initialized linear projection layer is added on top of contextual transformer network $g(\cdot)$ into $V$ classes representing the vocabulary entries of the task considered. The model then is fine-tuned with standard Connectionist Temporal Classification (CTC) loss \cite{graves2006connectionist} in a supervised fashion.
\vspace{-0.4cm}
\subsection{Speech Feature Extraction}
\vspace{-0.2cm}
The SSL pre-trained wav2vec2.0 model can be used as a standalone speech recognition system after task fine-tuning. Alternatively their contextual speech representations can be incorporated into various back-end target domain data trained ASR systems via feature fusion based on, e.g. TDNN or Conformer architectures, as considered in this paper. To this end, a ``Bottleneck module'' (circled in blue, top middle in Fig. 1) is introduced immediately above the highest positioned contextual transformer block to produce more compact speech representations\footnote{Alternative approaches of extracting bottleneck features, e.g. either from the CNN feature encoder outputs, contextual transformer network outputs, or a weighted average of the hidden layer outputs of each transformer block as suggested in \cite{pepino2021emotion} led to performance degradation.}. The ``Bottleneck module'' contains a stack of four interleaving convolutional and feedforward layers: the first 1D transposed de-convolution CNN layer is used to change the stride length to 10ms to allow a frame rate synchronization with that of the back-end ASR systems; a fully connected (FC) block, which consists of a linear layer, rectified linear unit (ReLU) activation and dropout module, is used to change the dimension of extracted features from 1024 to 256\footnote{256 empirically outperforms other \#dim, e.g. 128, 512 or 768}. This is followed by a CNN layer and a final FC block which are used  to revert the stride back to 20ms and restore the dimensionality to 1024. The final wav2vec2.0 speech representations are extracted from the first FC layer outputs and concatenated with the standard acoustic front-ends via feature fusion in hybrid TDNN or Conformer E2E ASR systems ((a) or (b) in Fig. 1).
\vspace{-0.3cm}
\section{Wav2vec2.0 features based A2A inversion}
\vspace{-0.2cm}
In this paper, the out-of-domain non-aged normal speech of the TaL dataset \cite{ribeiro2021tal} containing parallel ultrasound, video data collected from 81 native speakers of English is used to construct mixture density networks (MDN) based neural A2A inversion models \cite{hu2022exploiting}. Due to the large acoustic domain mismatch, a direct cross-domain application of the A2A inversion model trained on normal, non-aged acoustic-articulatory parallel data to the elderly and dysarthric speech is problematic, as shown in the previous researches on cross-domain A2A inversion \cite{hu2022exploiting,hu2022exploitinguti}. To this end, the large acoustic domain mismatch is minimized using either baseline multi-level adaptive networks (MLAN) \cite{liu2021recent, hu2022exploiting, hu2022exploitinguti} or the above domain fine-tuned SSL wav2vec2.0 speech representations of Section 2.3.
\par
The SSL based domain adaptation includes the following steps: 1) The pre-trained wav2vec2.0 model is fine-tuned on both out-of-domain normal speech and in-domain dysarthric or elderly speech simultaneously; 2) the resulting fine-tuned model is then used to produce more domain invariant speech representations that will exhibit smaller mismatch between these types of data. The resulting domain adapted speech representations are used in A2A inversion model training and articulatory feature generation for dysarthric or elderly audio data. More details of the MDN based A2A inversion models can be found in \cite{hu2022exploiting}. The articulatory movement features generated from dysarthric or elderly speech are concatenated with the standard acoustic features via feature fusion to construct multi-modal TDNN or Conformer systems ((c) or (d) in Fig. 1).
\vspace{-0.3cm}
\section{integration with TDNN/Conformer system}
\vspace{-0.2cm}
In order to exploit the diversity and complementarity among the SSL pre-trained wav2vec2.0 model and back-end TDNN or Conformer ASR systems, and to improve the generalization performance on unseen words as well as dysarthric speech of very low intelligibility, a series of integration approaches are explored in this paper.
\par
\noindent
\textbf{1) input feature fusion} is used to fuse standard acoustic frontends and domain adapted wav2vec2.0 speech representations via either feature concatenation before fed into the TDNN/Conformer system input layer, {or optionally feature fusion at the TDNN hidden layer (connections (a) and (b) in Fig. 1, centre left).
\par
\noindent
\textbf{2) time-synchronous frame-level joint decoding} is used to combine two or more hybrid TDNN systems separately trained using either standard acoustic features alone, or with additional domain fine-tuned wav2vec2.0 representations, by a frame level linear interpolation of system specific acoustic scores (green dotted box (e) in Fig. 1, bottom right). Let $P_{a+w}(y_t|X_1^{1:T})$ and $ P_{a}(y_t|X_2^{1:T})$ denote the frame level acoustic scores at $t$-th frame of TDNN systems constructed with or without additional wav2vec2.0 representations. Then the final frame-level score is obtained by
\begin{small}
\begin{equation}
    \vspace{-0.1cm}
    \setlength{\abovedisplayskip}{2pt}
    \setlength{\belowdisplayskip}{2pt}
    {\hat P(y_t|X_1^{1:T}, X_2^{1:T})} = \alpha P_{a+w}(y_t|X_1^{1:T}) + \beta P_{a}(y_t|X_2^{1:T})
\end{equation}
\end{small}
\par
\noindent
where $\alpha$ and $\beta$ are the weights assigned to each system.
\par
\noindent
\textbf{3) cross system multi-pass decoding} involves a first decoding pass TDNN or Conformer system produced N-best outputs to be rescored using domain adapted wav2vec2.0 models.  Consider a sequence of acoustic features $X$ and its corresponding TDNN or Conformer (cfm) decoded N-best recognition hypotheses in \{$ Y_1, Y_2, ..., Y_n$\}. Let ${\textbf{s}}^{w2v} = [s_1^{w2v}, s_2^{w2v}, ..., s_n^{w2v}]^T$ and ${\textbf{s}}^{tdnn/cfm} = [s_1^{tdnn/cfm}, s_2^{tdnn/cfm}, ..., s_n^{tdnn/cfm}]^T$ denote the N-best score vectors produced by the first pass TDNN/Conformer and the second pass wav2vec2.0 models. The final 1-best output produced in the second pass wav2vec2.0 rescoring is obtained by interpolating the system specific N-best entry level scores as 
\begin{small}
\begin{equation}
    \setlength{\abovedisplayskip}{2pt}
    \setlength{\belowdisplayskip}{2pt}
    {\hat Y}_{best} = \mathop{\arg}\min_{i} \{\alpha s_i^{w2v} + \beta s_i^{tdnn/cfm}\}
    \vspace{-0.1cm}
\end{equation}
\end{small}
\par
\noindent
where $\alpha$ and $\beta$ are the weights assigned to the second pass wav2vec2.0 and the first pass TDNN or Conformer systems. 
An example of two-pass rescoring based TDNN/Conformer and wav2vec2.0 system combination is shown in Fig. 1 (top right, red box (f)).

\begin{table*}[htbp]
\centering
\caption{The performance of wav2vec2.0 based ASR,  TDNN/Conformer based systems constructed without or with wav2vec2.0 (w2v) representations and optionally using the cross-domain inverted UTI articulatory features on 1) the UASpeech test set; AND 2) the DementiaBank Pitt development (Dev) and evaluation (Eval) sets; ``VL/L/M/H'' refer to intelligibility subgroups. ``INV'' and ``PAR'' refer to clinical investigator and elderly participant. ``ada. w2v'' and ``ada. BN'' are the abbreviations of cross-domain adapted wav2vec2.0 representations and bottleneck features from MLAN\cite{hu2022exploitinguti} respectively. ``+'' and ``$\rightarrow$'' stand for frame-level joint decoding and multi-pass decoding. $\dag$ and $\ast$ denote statistical significant (MAPSSWE,$\alpha$ = 0.05) differences obtained against the baseline wav2vec2.0 and Conformer systems (Sys. 4,12).}
\scalebox{0.7}{
\begin{tabular}{c|c|c|c|cc|cccc|c|cc|cc|c} 
\hline\hline
\multirow{3}{*}{Sys} & \multirow{3}{*}{model}  & \multirow{3}{*}{\begin{tabular}[c]{@{}c@{}}acoustic \\ feature\end{tabular}} & \multirow{3}{*}{A2A input} & \multicolumn{7}{c|}{UASpeech WER (\%)}                                                                                                                                  & \multicolumn{5}{c}{DBank WER(\%)}                                                                                      \\ 
\cline{5-16}
                     &                         &                                                                              &                            & \multirow{2}{*}{unseen} & \multirow{2}{*}{seen} & \multirow{2}{*}{VL}   & \multirow{2}{*}{L}    & \multirow{2}{*}{M}    & \multirow{2}{*}{H}    & \multirow{2}{*}{All} & \multicolumn{2}{c|}{Dev.}                     & \multicolumn{2}{c|}{Eval.}                    & \multirow{2}{*}{All}  \\ 
\cline{12-15}
                     &                         &                                                                              &                            &                         &                       &                       &                       &                       &                       &                       & PAR.                  & INV.                  & PAR.                  & INV.                  &                        \\ 
\hline\hline
$1$                  & \multirow{2}{*}{TDNN-f} & \multirow{2}{*}{FBK}                                                         & \xmark                            & 51.62                   & 16.98                 & 62.53                 & 31.92                 & 23.12                 & 13.67                 & 30.56                 & 47.93                 & 19.91                 & 36.66                 & 19.76                 & 33.80                  \\
$2$                  &                         &                                                                              & ada. BN                    & 54.95                   & 17.64                 & 63.35                 & 33.14                 & 25.76                 & 15.78                 & 32.27                 & 45.82                 & 19.21                 & 34.89                 & 18.42                 & 32.35                  \\ 
\hline
$3$                  & +LHUC                   & FBK                                                                          &   \xmark                         & 47.73                   & 15.49                 & 61.12                 & 28.95                 & 19.08                 & 11.94                 & 28.13                 & 45.49                 & 19.26                 & 35.44                 & 18.42                 & 32.33                  \\ 
\hline\hline
$4$                  & wav2vec2                & raw audio                                                                    &  \xmark                          & 55.75                   & 14.67                 & 62.76                 & 36.91                 & 25.24                 & 9.11                  & 30.78                 & 29.71                 & 14.29                 & 21.27                 & 15.32                 & 21.60                  \\ 
\hline
$5$                  & \multirow{2}{*}{TDNN-f} & \multirow{2}{*}{FBK+w2v}                                                     & \xmark                            & 55.01          & 14.65                 & 60.59$^\dag$ & 34.81$^\dag$ & 23.80$^\dag$ & 12.01                 & 30.48        & 27.21$^\dag$ & 13.20$^\dag$ & 19.13$^\dag$ & 12.87                 & 19.73$^\dag$  \\
$6$                  &                         &                                                                              & ada. w2v                   & \textbf{53.90$^\dag$}   & \textbf{14.17}        & \textbf{60.07$^\dag$} & \textbf{33.58$^\dag$} & \textbf{23.49$^\dag$} & 11.33                 & \textbf{29.75$^\dag$} & \textbf{26.43$^\dag$} & \textbf{12.77$^\dag$} & \textbf{18.29$^\dag$} & \textbf{12.10$^\dag$} & \textbf{19.08$^\dag$}  \\ 
\hline
$7$                  & Sys.3 + 5               & -                                                                            & -                          & 43.77$^\dag$   & 11.72$^\dag$ & 53.19$^\dag$ & 26.33$^\dag$ & 16.47$^\dag$ & 9.02                  & 24.29$^\dag$ & 26.98$^\dag$ & 12.82$^\dag$ & 18.06$^\dag$ & 12.21                 & 19.29$^\dag$  \\
$8$                  & Sys.3 + 6               & -                                                                            & -                          & 43.68$^\dag$   & 11.71$^\dag$ & 53.39$^\dag$ & 26.17$^\dag$ & 16.57$^\dag$ & 8.84                  & 24.25$^\dag$ & 26.11$^\dag$ & 12.60$^\dag$ & 17.80$^\dag$ & 11.32$^\dag$ & 18.78$^\dag$  \\
$9$                  & Sys.3+5+6               & -                                                                            & -                          & \textbf{43.54$^\dag$}   & \textbf{11.68$^\dag$} & \textbf{53.26$^\dag$} & \textbf{26.13$^\dag$} & \textbf{16.45$^\dag$} & 8.81                  & \textbf{24.18$^\dag$} & \textbf{26.05$^\dag$} & \textbf{12.52$^\dag$} & \textbf{17.59$^\dag$} & \textbf{11.43$^\dag$} & \textbf{18.69$^\dag$}  \\ 
\hline
$10$                 & Sys.3$\rightarrow$4     & -                                                                            & -                          & \textbf{40.06$^\dag$}   & \textbf{11.72$^\dag$} & \textbf{52.53$^\dag$} & \textbf{25.00$^\dag$} & \textbf{14.80$^\dag$} & \textbf{7.08$^\dag$}  & \textbf{22.83$^\dag$} & 34.17                 & 15.41                 & 25.49                 & 15.98                 & 24.55                  \\
$11$                 & Sys.9$\rightarrow$4     & -                                                                            & -                          & \textbf{39.09$^\dag$}   & \textbf{11.90$^\dag$} & \textbf{53.12$^\dag$} & \textbf{25.03$^\dag$} & \textbf{14.04$^\dag$} & \textbf{6.32$^\dag$}  & \textbf{22.56$^\dag$} & \textbf{25.27$^\dag$} & \textbf{12.28$^\dag$} & \textbf{16.84$^\dag$} & \textbf{11.54$^\dag$} & \textbf{18.17$^\dag$}  \\ 
\hline\hline
$12$                 & \multirow{2}{*}{Conformer}               & FBK                                                                          & \xmark                           & 99.30                   & 18.24                 & 66.77                 & 49.39                 & 46.47                 & 42.02                 & 50.03                 & 48.71                 & 20.97                 & 36.93                 & 19.42                 & 34.57                  \\ 
$13$                 &               & FBK+w2v                                                                      & ada. w2v                   & \textbf{72.80$^\ast$}   & \textbf{14.87$^\ast$} & \textbf{64.24$^\ast$} & \textbf{43.09$^\ast$} & \textbf{36.41$^\ast$} & \textbf{17.28$^\ast$} & \textbf{37.59$^\ast$} & \textbf{28.38$^\ast$} & \textbf{14.53$^\ast$} & \textbf{19.40$^\ast$} & \textbf{13.10$^\ast$} & \textbf{20.79$^\ast$}  \\ 
\hline
$14$                 & Sys.13$\rightarrow$4    & -                                                                            & -                          & \textbf{71.73$^\ast$}   & \textbf{14.50$^\ast$} & \textbf{63.46$^\ast$} & \textbf{42.75$^\ast$} & \textbf{35.69$^\ast$} & \textbf{16.53$^\ast$} & \textbf{36.95$^\ast$} & \textbf{27.89$^\ast$} & \textbf{14.28$^\ast$} & \textbf{19.15$^\ast$} & \textbf{13.10$^\ast$} & \textbf{20.45$^\ast$}  \\
\hline\hline
\end{tabular}
}
\vspace{-0.6cm}
\end{table*}
\vspace{-0.3cm}
\section{Experiments}
\vspace{-0.2cm}
\noindent
\textbf{Task Description:} The UASpeech corpus is the largest publicly available dysarthric speech corpus \cite{kim2008dysarthric} consisting of 16 dysarthric and 13 control speakers, and contains 155 common words and 300 uncommon words. The entire corpus is further divided into 3 subset blocks per speaker. The same set of 155 common words is used in all three blocks, while the uncommon words in each block are different. The data from Block 1 and 3 of all the 29 speakers are used as the training set, while the data of Block 2 collected from all the 16 dysarthric speakers serves as the test set. After removing excessive silence and speed perturbation based data augmentation \cite{geng2020investigation}, a total of 130.1 hours of audio data is used as the training set, while 9 hours of speech is used for performance evaluation. The DementiaBank Pitt \cite{becker1994natural} corpus contains about 33-hour audio recorded over interviews between the 292 elderly participants and clinical investigators. 
After silence stripping and data augmentation, the duration of training data is increased to 58.9 hours, while the development and evaluation sets contain 2.5 hours and 0.6 hours of audio respectively. The test data word coverage rates of UASpeech and DementiaBank are 61\% and 98.7\%.
\\
\noindent
\textbf{Experimental Setup:} The pre-trained wav2vec2.0 models on UASpeech and DementiaBank corpora are the Large Wav2Vec2.0 model\footnote{https://huggingface.co/facebook/wav2vec2-large-960h-lv60} and Conformer based Wav2Vec2.0 with relative position embeddings\footnote{https://huggingface.co/facebook/wav2vec2-conformer-rel-pos-large-960h-ft} respectively which are pretrained with 60k hours of Libri-light data and fine-tuned with 960 hours of Librispeech data. During domain adaptation to UASpeech data, the wav2vec2.0 model is first fine-tuned with the non-augmented data to update all parameters except those of speech feature encoder, then further fine-tuned with the 130.1h augmented data to only update the parameters of the first 12 transformer blocks to ensure generalization to unseen words. The numbers of training epochs for the two fine-tuning stages are both 20. For the DementiaBank data, all the wav2vec2.0 parameters except those of the speech feature encoder are fine-tuned for 30 epochs on the 58.9h augmented data. The initial learning rate is 1e-5 and then linearly decays. As section 2.3 mentioned, the dimensionality of bottleneck wav2vec2.0 speech representations is 256 and that of ultrasound based articulatory features is 144. Experiment setups of back-end TDNN and Conformer systems follow \cite{jin2022personalized}. \\
\noindent
\textbf{Results on UASpeech:} \textbf{1)} 
Although there is no significant overall WER difference between the baseline TDNN with standard 40-dim filter-bank (FBK) features and the domain fine-tuned wav2vec2.0 model in Table 1 (Sys.1 vs. 4), the wav2vec2.0 model degrades the performance by 4.13\% absolute on the unseen subset. 
\textbf{2)} Fusing FBK front-ends with the fine-tuned wav2vec2.0 features (Fig. 1(a)-(b)) as TDNN system inputs (Sys. 5), or optionally further incorporating cross-domain A2A inverted features (Fig. 1(c)-(d)) for multimodal TDNN system training (Sys. 6), consistently reduce the overall WER by 0.3\% and 1.03\% absolute over the wav2vec2.0 model (Sys.5, 6  vs. 4). \textbf{3)} Large overall WER reductions of 6.49-6.60\% absolute (11.98-12.21\% on unseen words) over the  wav2vec2.0 model are obtained by frame-level joint decoding between the LHUC-SAT TDNN system trained using FBK features alone, and those trained with additional wav2vec2.0 representations and inverted articulatory features (Sys.7-9 vs. 4)\footnote{System weights empirically set as 3:2, 3:2 and 9:1:5 for Sys. 7, 8 and 9}. \textbf{4)} Further performance improvements are obtained by cross system multi-pass decoding (Sys.10, 11), which uses the wav2vec2.0 model to rescore the joint decoded TDNN systems outputs. The lowest published WER of 22.56\% on 16 dysarthric speakers (Sys.10-11, 52.53\% on very low intelligibility, 39.09\% on unseen words) is obtained\footnote{System weights empirically set as 2:9 and 4:15 for Sys.10 and 11}. \textbf{5)} Similar experiments on Conformer systems also demonstrate the effectiveness of the wav2vec2.0 features via feature fusion (Sys.13 vs. 12), where smaller system combination improvements are obtained by multi-pass decoding (Sys.14 vs. 13), due to architecture similarity between the wav2vec2.0 and Conformer models leading to reduced cross system complementarity. Finally, the performance of our best systems (Sys.10, 11, Table 1) are contrasted against published state-of-the-art results on UASpeech in Table 2. 

\begin{table}[htbp]
\vspace{-0.6cm}
\centering
\caption{WER (\%) of published and our best systems on UASpeech. ``15 spk'' stands for only 15 speakers in the test set, but the same 4 speakers are used in ``very low'' (VL) intelligibility group.}
\scalebox{0.72}{
\begin{tabular}{c|c|c} 
\hline\hline
System                                                                   & VL      & All     \\ 
\hline\hline
Sheffield-2020 Fine-tuning CNN-TDNN speaker adaptation \cite{xiong2020source} (15 spk)                & 68.24 & 30.76  \\
CUHK-2021 NAS  DNN  +  Data  Aug.  +  LHUC-SAT  +  AV  fusion \cite{liu2021recent}        & 60.30 & 25.21  \\
CUHK-2022 DNN + Data Aug. + LHUC-SAT + AUV fusion \cite{hu2022exploiting}                    & 60.14 & 24.82  \\
CUHK-2022 DNN + Data Aug. + SBE Adapt + LHUC-SAT \cite{geng2022speaker}                             & 59.30 & 25.05  \\
CUHK-2022 TDNN + spectral basic GAN + LHUC-SAT \cite{jin2022personalized}                        & 59.18 & 27.85  \\
BUT-2022 Wav2vec2 + fMLLR + xvectors \cite{baskar2022speaker} (15 spk)                                 & 57.72 & 22.83  \\
\textbf{TDNN + wav2vec2 fea. + sys. combination (sys. 10 Table 1, ours) }& \textbf{52.53} & 22.83  \\
\textbf{TDNN + wav2vec2 fea. + sys. combination (sys. 11 Table 1, ours)} & 53.12 & \textbf{22.56}  \\
\hline\hline
\end{tabular}
}
\vspace{-0.4cm}
\end{table}
\noindent
\textbf{Results on DementiaBank:} \textbf{1)} Compared with the baseline TDNN system trained with 40-dim FBK features, the domain fine-tuned wav2vec2.0 model produces consistent and large WER reduction of 12.20\% absolute  (Sys.4 vs. 1). \textbf{2)} Similar to the UASpeech experiments, the incorporation of wav2vec2.0 features alone, or plus their corresponding cross-domain articulatory features, produces significant WER reduction up to 2.52\% (Sys.6 vs. 4). \textbf{3)} Frame-level joint decoding between the LHUC-SAT TDNN system using FBK features (Sys. 3), and the TDNN systems using additional wav2vec2.0 features (Sys. 5) and articulatory features (sys. 6) consistently outperforms the adapted wav2vec2.0 model (sys. 4) by a significant WER reduction of 2.91\% absolute (13.47\% relative, Sys. 9 vs. 4). \textbf{4)} After the cross system multi-pass decoding, the lowest published WER of 18.17\% on DementiaBank Pitt test set is obtained (Sys.11). Using its final recognition outputs on evaluation set to extract textual features, a state-of-the-art speech recognition based AD detection \cite{wang22l_interspeech} accuracy of 91.7\% is also obtained. A comparison of WERs and AD detection accuracy on DementiaBank between previously published ASR systems and our system is shown in Table 3.
\vspace{-0.6cm}
\begin{table}[htbp]
\centering
\caption{WER (\%) and AD detection accuracy (\%) between published systems on DementiaBank and our system.}
\scalebox{0.66}{
\begin{tabular}{c|c|c|c} 
\hline\hline
\multirow{2}{*}{System}                                         & \multicolumn{2}{c|}{WER (\%)} & \multirow{2}{*}{Acc}  \\ 
\cline{2-3}
& All    & Eval PAR.      &                       \\ 
\hline\hline
TDNN + Data Aug. + domain adapt. + LHUC-SAT \cite{ye2021development}                           & 29.90 & 33.17        & 87.5                \\
TDNN + Data Aug. + SBE Adapt + LHUC-SAT \cite{geng2022speaker}                                & 29.16 & 31.89        & -                     \\
TDNN + spectral basic GAN + LHUC-SAT \cite{jin2022personalized}                           & 31.60 & 34.08        & -                     \\
Conformer + NAS + domain \& spk adapt  + sys. combination \cite{wang2022conformer}        & 24.2  & 25.5         & 87.5                \\
\textbf{TDNN + wav2vec2 fea. + sys. combination (sys. 11 table 1, ours)} & \textbf{18.17} & \textbf{16.84}        & \textbf{91.7}                \\
\hline\hline
\end{tabular}
}
\end{table}
\vspace{-0.7cm}
\section{CONCLUSION}
\vspace{-0.3cm}

This paper explores a series of approaches to integrate cross-domain adapted SSL pre-trained models into TDNN and Conformer ASR systems for dysarthric and elderly speech including input feature fusion, frame-level joint decoding and multi-pass decoding. In addition, domain adapted wav2vec2.0 representations are utilized in acoustic-to-articulatory (A2A) inversion to construct multi-modal dysarthric and elderly ASR systems. The TDNN and Conformer ASR systems integrated domain adapted wav2vec2.0 models consistently outperform the standalone domain fine-tuned wav2vec2.0 models by statistically significant WER reductions of 8.22\% and 3.43\% absolute (26.71\% and  15.88\% relative) on the benchmark UASpeech and DementiaBank Pitt test set respectively, while producing the lowest published WERs of 22.56\% and 18.17\% on the two tasks. Future research will study the personalization of pre-trained ASR models for dysarthric and elderly speakers and pre-trained ASR model compression approaches to make it more practically applicable.
\vspace{-0.5cm}
\section{Acknowledgement}
\vspace{-0.2cm}
This research is supported by Hong Kong RGC GRF grant No. 14200220, 14200021, TRS T45-407/19N, Innovation \& Technology Fund grant No. ITS/218/21, 
and National Natural Science Foundation of China (NSFC) Grant 62106255.
\clearpage
\bibliographystyle{IEEEtran}
\bibliography{refs}
\end{document}